\documentclass[usenatbib]{mn2e}
\usepackage{graphicx}
\usepackage{ifthen}
\usepackage{url}

\def\ltsima{$\; \buildrel < \over \sim \;$}
\def\simlt{\lower.5ex\hbox{\ltsima}}
\def\gtsima{$\; \buildrel > \over \sim \;$}
\def\simgt{\lower.5ex\hbox{\gtsima}}
\def\kpc{{\rm\,kpc}}

\def\msun{{\rm\,M_\odot}}
\def\lsun{{\rm\,L_\odot}}

\def\pc{{\rm\,pc}}

\newcommand{\fmmm}[1]{\mbox{$#1$}}
\newcommand{\scnd}{\mbox{\fmmm{''}\hskip-0.3em .}}

\newcommand{\mcnd}{\mbox{\fmmm{'}\hskip-0.3em .}}

\def\deg{^\circ}

\def\s{\ifmmode \widetilde \else \~\fi}
\def\={\overline}

\def\spose#1{\hbox to 0pt{#1\hss}}

\def\lta{\mathrel{\spose{\lower 3pt\hbox{$\mathchar"218$}}
     \raise 2.0pt\hbox{$\mathchar"13C$}}}
\def\gta{\mathrel{\spose{\lower 3pt\hbox{$\mathchar"218$}}
     \raise 2.0pt\hbox{$\mathchar"13E$}}}
\def\Dt{\spose{\raise 1.5ex\hbox{\hskip3pt$\mathchar"201$}}}    
\def\dt{\spose{\raise 1.0ex\hbox{\hskip2pt$\mathchar"201$}}}    

\def\dotsfill{\leaders\hbox to 1em{\hss.\hss}\hfill}

\def\FeH{{\rm[Fe/H]}}

\loadboldmathitalic 
\title[Three new faint dwarfs in the outer halo of the Andromeda galaxy]{Discovery and analysis of three faint
dwarf galaxies and a globular cluster in the outer halo of the Andromeda galaxy$^8$}
\author[N. F. Martin et al.] {N. F. Martin$^{1}$, R. A. Ibata$^{1}$, M. J. Irwin$^{2}$, S. Chapman$^{3}$, G. F.
Lewis$^{4}$,
\newauthor A. M. N. Ferguson$^{5}$, N. Tanvir$^{6}$ \& A. W. McConnachie$^{7}$\\
$^{1}$ Observatoire de Strasbourg, 11, rue de l'Universit\'e, F-67000, Strasbourg, France\\
$^{2}$ Institute of Astronomy, Madingley Road, Cambridge, CB3 0HA, U.K.\\
$^{3}$ California Institute of Technology, Pasadena, CA, 91125, USA \\
$^{4}$ Institute of Astronomy, School of Physics, A29, University of Sydney, NSW 2006, Australia\\
$^{5}$ Institute for Astronomy, University of Edinburgh, Royal Observatory, Blackford Hill, Edinburgh, EH9 3HJ, U.K.\\
$^{6}$ Physical Sciences, Univ. of Hertfordshire, Hatfield, AL10 9AB, U.K.\\
$^{7}$ Department of Physics and Astronomy, University of Victoria, Victoria, B.C., V8P 1A1, Canada}

\date{\today}
\begin{document} 
\maketitle 
\begin{abstract} 
We present the discovery of three faint dwarf galaxies and a globular cluster in the halo of the Andromeda galaxy (M31),
found in our MegaCam survey that spans the southern quadrant of M31, from a projected distance of $\sim50\kpc$ to
$\sim150\kpc$. Though the survey covers 57~sq.~degrees, the four satellites lie within $2\deg$ of one another.
From the tip of the red giant branch, we estimate that the globular cluster lies at a distance of $631\pm58\kpc$ from
the Milky Way and along with a $\sim100\kpc$ projected distance from M31 we derive a total distance of $175\pm55\kpc$
from its host, making it the farthest M31 globular cluster known. It also shows the typical characteristics of a bright
globular cluster, with a half-light radius of $2.3\pm0.2\pc$ and an absolute magnitude in the $V$ band of
$M_{V,0}=-8.5\pm0.3$. Isochrone fitting reveals it is dominated by a very old population with a metallicity of
$\FeH\sim-1.3$. The three dwarf galaxies are revealed as overdensities of stars that are aligned along red giant
branch tracks in their color-magnitude diagrams. These satellites are all very faint, with absolute magnitudes in the
range $-7.3\simlt M_{V,0} \simlt -6.4$, and show strikingly similar characteristics with metallicities of
$\FeH\sim-1.4$ and half-light radii of $\sim120\pm45\pc$, making these dwarf galaxies two to three times smaller than
the smallest previously known satellites of M31. Given their faintness, their distance is difficult to constrain, but we
estimate them to be between 740 and $955\kpc$ which places them well within the virial radius of the host galaxy. The
panoramic view of the MegaCam survey can provide an unbiased view of the satellite distribution of the Andromeda galaxy
and, extrapolating from its coverage of the halo, we estimate that up to $45\pm20$ satellites brighter than
$M_V\sim-6.5$ should be orbiting M31. Hence faint dwarf galaxies cannot alone account for the missing satellites that
are predicted by $\Lambda$CDM models, unless they reside in dark matter mini-halos that are more massive than the
typical masse of $10^7\msun$ currently inferred from their central radial velocity dispersion.
\end{abstract}
 

\section{Introduction}
\footnotetext[8]{Based on observations obtained with MegaPrime/MegaCam, a joint project of CFHT and CEA/DAPNIA, at
the Canada-France-Hawaii Telescope (CFHT) which is operated by the National Research Council (NRC) of Canada, the
Institut National des Science de l'Univers of the Centre National de la Recherche Scientifique (CNRS) of France, and the
University of Hawaii.}

A generic prediction of Cold Dark Matter (CDM) cosmology is the existence of copious sub-structure in gravitationally
collapsed structures such as galaxy clusters, galaxy halos and even dwarf galaxies \citep{klypin99,moore99a,moore01}. On
the scale of large spiral galaxies like the Milky Way and the Andromeda galaxy (M31), upwards of 500 dense dwarf galaxy-mass  
clumps are expected to orbit in the halo. These structures have very dense cores, with a very steep radial profile
\citep{navarro97,moore99b}, which renders them essentially impervious to Galactic tides. However, more than two orders
of magnitude fewer luminous dwarf galaxies have so far been found in the Local Group.

Several studies have investigated the effect of reionization on the early evolution of small structures
\citep[e.g.][]{bullock00,sommerville02,tully02}. Any gas that is not in deep potential wells is lost from the
protogalaxies as the first ionizing sources turn on. This ``squelching'' of star-formation could solve the CDM satellite
over-production problem by rendering low-mass galaxies invisible, or at least very dark, depending on the gas fraction
that managed to cool to dense molecular form before reionization. Generally, the observed distribution of dwarf
satellites can be brought into agreement with CDM if dwarf galaxies reside in large dark matter halos \citep{stoehr02},
but it opens up the possibility that lower mass dwarfs are being missed.

The recent discoveries of faint dwarf galaxies around both the Andromeda galaxy and our own Milky Way could be the first
step in uncovering such satellites. Andromeda~IX (And~IX, \citealt{zucker04,chapman05,harbeck05}), 
Andromeda~X (And~X, \citealt{zucker06a}), Ursa Major (UMa, \citealt{willman05}), Canes Venatici (CVn,
\citealt{zucker06b}) and Bootes (Boo, \citealt{belokurov06}) are all fainter than $M_V\sim-8.3$ but show characteristics
typical of dwarf galaxies. Radial velocity measurements show that at least And~IX \citep{chapman05}, UMa
\citep{kleyna05} and Boo \citep{munoz06} appear to be highly dark matter dominated and could correspond to the dark
matter sub-structures that are found in the CDM models. The discovery of all these faint dwarf galaxies within
wide-field surveys (the Sloan Digital Sky Survey and the INT Wide Field Camera survey of the inner halo of M31;
\citealt{ibata01} and \citealt{ferguson02}) revealed the need to conduct systematic surveys of large regions of the halos
of the Milky Way or the Andromeda galaxy in order to properly constrain the low luminosity end of their satellite
distribution with adequate statistics.

Even though the M31 galaxy is located at $785\pm 25\kpc$ \citep{mcconnachie05} from the Sun, surveying the Andromeda
galaxy halo has the advantage of requiring to map a much smaller area to have a panoramic view of its halo compared to
the Milky Way within which we sit and whose disk and bulge significantly hamper observations at low latitude. Hence we
undertook a systematic survey of the outer part of the M31 halo that extends the inner halo survey conducted by our
team, which has revealed a flurry of substructures \citep{ibata01,ferguson02,mcconnachie04,ibata05}. The new survey was
conducted with the wide field camera MegaCam mounted on the Canada-France-Hawaii Telescope and spans the whole Southern
quadrant of the M31 halo. Covering 57 square degrees, it extends from a projected distance of $\sim 50$ to $\sim
150\kpc$ at the distance of M31.

In this paper, we present the discovery of the three new faint dwarf galaxies that appear in this survey, revealed by
an overdensity of stars on the sky that follow red giant branch (RGB) tracks at the distance of M31. The farthest
globular cluster in the halo of M31 is also reported. The dataset is presented in section~2 and the characteristics of
the four new satellites are derived in section~3 and discussed in section~4. We conclude in section~5. We refer the
reader to a companion paper \citep{ibata06} for the search and analysis of large scale structures in the MegaCam survey.

\section{Observations and reduction}
The MegaCam instrument is a wide field camera composed of a mosaic of 36 CCDs that cover a total area of
$0.96\deg\times0.94\deg$. Each CCD is composed of $2048 \times 4612$ pixels of 0.187 arcseconds. The camera is mounted
on the 3.6\,m Canada-France-Hawaii Telescope on Mauna Kea, ensuring excellent image quality. Each of the 63 fields
of the survey was observed in the $g$ and $i$ bands with an exposure time of 1450\,s per filter in seeing better than
$0\mcnd8$. This yielded limiting magnitudes of $\sim25.5$ and $\sim24.5$ in the $g$ and $i$ bands, respectively. 

The raw images were pre-processed by the CFHT team using the Elixir
system\footnote{\url{http://www.cfht.hawaii.edu/Instruments/Imaging/MegaPrime/dataprocessing.html}} in order to correct
for the instrumental signature across the whole mosaic. Then we reduced the data using a version of the CASU pipeline
\citep{irwin01} adapted for MegaCam observations to produce catalogues of magnitudes, colors and object morphological
classifications. The $g$ and $i$ magnitudes are de-reddened using the $E(B-V)$ values from \citet{schlegel98} IRAS
maps as follows: $g_0=g-3.793\cdot E(B-V)$ and $i_0=i-2.086\cdot E(B-V)$ where $g_0$ and $i_0$ are the de-reddened
magnitudes. The extinction values however remain low ($E(B-V)<0.1$) and do not vary over the small regions of a few
arcminutes considered here (see also \citealt{ibata06}).

\begin{figure}
\begin{center}
\includegraphics[width=0.7\hsize, angle=270]{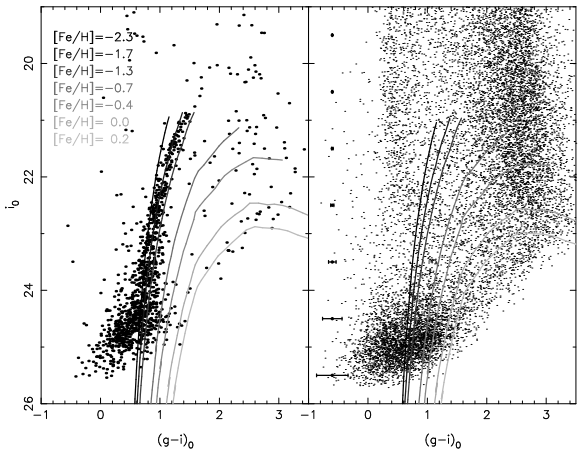}
\caption{\emph{Left:} CMD of a region of $3'$ around the Andromeda~III dwarf galaxy. \citet{girardi04} isochrones have
been overlaid for a population of 14.15~Gyr with $\FeH=-2.3$, $-1.7$, $-1.3$, $-0.7$, $-0.4$, $0.0$ and $+0.2$ (from
left to right) and at the distance of And~III ($m-M=24.37$). \emph{Right:} CMD of a one square degree MegaCam field in
the outer parts of the M31 halo. The isochrones are also overlaid and correspond to regions that are not significantly
contaminated by Galactic disk stars ($2.0\simlt (g-i)_0 \simlt3.0$, $i_0\simlt23.0$) and halo stars ($0.2\simlt
(g-i)_0\simlt0.8$ and $i_0\simlt22.0$). The mean uncertainty on the $i$ magnitude values are shown as a function of
magnitude on the left of the panel.}
\label{M31dwarf_fig01}
\end{center}
\end{figure}

Comparing the observed RGB stars of the new satellites with isochrones is essential for deriving their parameters. We
use the \citet{girardi04} isochrones for comparison since they are available in the SDSS $g$ and $i$ system. To avoid
age-metallicity degeneracy, we choose to use only old isochrones (14~Gyr) as halo globular clusters and the dominant
dwarf galaxy stellar populations are generally found to be old (see e.g. \citealt{mackey04} for Galactic halo globular
clusters and \citealt{grebel01} for dwarf galaxies). To check the applicability of these isochrones we first compared
them to the Andromeda~III dwarf galaxy. The color-magnitude diagram (CMD) of stars within $3'$ of the center of the
dwarf galaxy is shown on the left panel of Figure~\ref{M31dwarf_fig01}. The \citet{girardi04} isochrones have been
overlaid on the CMD at the distance of the dwarf \citep[$m-M=24.37$, ][]{mcconnachie05} and are in good agreement with
the metallicity of And~III ($\FeH=-1.7$, \citealt{mcconnachie05}). In the following, an estimate of the metallicity of
each RGB star at the distance of M31 is provided by the ratio of the distances in the CMD from this star to the two
closest isochrones. Even though this will only be a crude estimation of the metallicity of this star, a more precise
determination is not possible given the uncertainties on the distance and age of the stars.

The right panel of Figure~\ref{M31dwarf_fig01} shows the CMD of a typical outer halo field of the MegaCam survey. The
isochrones are placed at the distance of M31 \citep[$m-M=24.47$; ][]{mcconnachie05} and reveal that the Galactic
contamination coming from disk stars at $2.0\simlt (g-i)_0 \simlt3.0$ and $i_0\simlt23.0$ and halo stars at $0.2\simlt
(g-i)_0\simlt0.8$ and $i_0\simlt22.0$ is mainly avoided, especially for the typical color of the stellar populations
found in the new satellites.

\section{Four new M31 companions}
The four new M31 companions were discovered directly during a visual inspection of the CMDs and sky maps of the 63
fields of the survey. They correspond to groups of stars that are at the same time clumped on the sky and aligned along
a RGB track. An automatic search to detect more diffuse substructures or those that could remain hidden in a denser
background population (e.g. in the giant stream that is also present in the survey region; \citealt{ibata01,ibata06})
will be presented in a companion paper (Martin et al., in prep). The four satellites described here are all recovered
with a high signal-over-noise ratio in the automatic search.

\begin{figure}
\begin{center}
\includegraphics[width=1.0\hsize, angle=270]{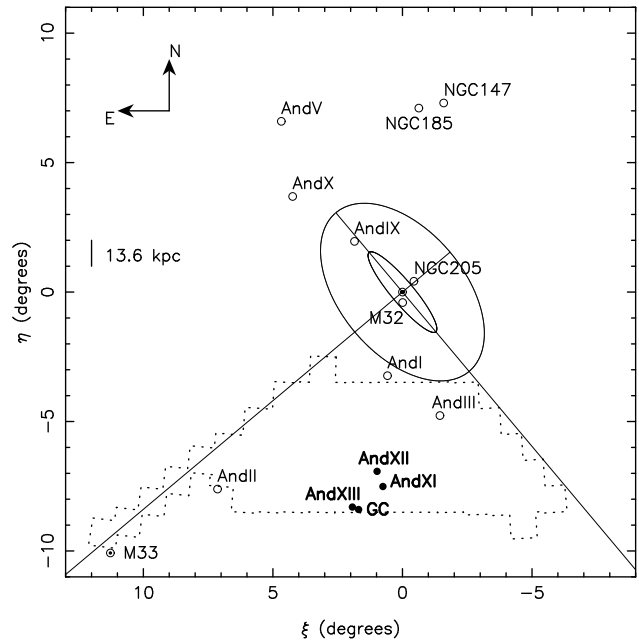}
\caption{Distribution of satellite dwarf galaxies around M31 (hollow circles; M31 is at the center of the map, North
is to the top and East to the left). At the distance of M31, $1\deg$ corresponds to $13.6\kpc$. The new MegaCam survey
is represented by the dotted line polygone in the southern quadrant and the four satellites discovered in the survey
(the three dwarf galaxies And~XI,~XII and~XIII and the globular cluster GC) are represented as filled circles. The inner
ellipse roughly represents the {\sc Hi} disk of Andromeda, with a radius of $27\kpc$ and the outer ellipse corresponds
to the inital coverage of the INT survey of \citet{ibata01,ibata05} and \citet{ferguson02}. It has since been extended
to the south and slightly overlaps with the MegaCam fields.}
\label{M31dwarf_fig02}
\end{center}
\end{figure}

Among the four satellites, one is a bright globular cluster (GC) and  three have the characteristics of faint dwarf
galaxies. Following the naming convention adopted by \citet{armandroff98} and \citet{armandroff99}, we
call these And XI, XII and XIII. Surprisingly, they all lie within two degrees of one another (see
Figure~\ref{M31dwarf_fig02}). The derived parameters of the four objects are summarized in Table~\ref{tableSat}.

\begin{table*}
\begin{center}
\caption{Properties of the four new M31 companions.}
\label{tableSat}
\begin{tabular}{l|ccc|c}
\hline\hline
 & And~XI & And~XII & And~XIII & GC\\
\hline
$\alpha$ (J2000) & $0^{\mathrm{h}}46^{\mathrm{m}}20^{\mathrm{s}}$ & $0^{\mathrm{h}}47^{\mathrm{m}}27^{\mathrm{s}}$ &
$0^{\mathrm{h}}51^{\mathrm{m}}51{\mathrm{s}}$ & $0^{\mathrm{h}}50^{\mathrm{m}}42^{\mathrm{s}}.5$\\
$\delta$ (J2000) & $+33\deg48'05''$ & $+34\deg22'29''$ & $+33\deg00'16''$ & $+32\deg54'59\scnd6$\\
$M_{V,0}^{\mathrm{min}}$ & $-5.0$ & $-4.3$ &$-4.6$ & -\\
$M_{V,0}$ & $-7.3\pm0.5$ & $-6.4\pm1.0$ &$-6.9\pm1.0$ & $-8.5\pm0.3$\\
$r_{1/2}$ & $0\mcnd5\pm0.2$ & $0\mcnd55\pm0.2$ & $0\mcnd5\pm0.2$ & $0\scnd76$\\
$r_{1/2}$ (pc) & $115\pm45$ & $125\pm45$ & $115\pm45$ & $2.3\pm0.1$\\
median $\FeH$  & $-1.3$ & $-1.5$ & $-1.4$ & $-1.3^{\mathrm{b}}$\\
$(m-M)_0$  & 24.47 (M31)$^{\mathrm{a}}$ & 24.47 (M31)$^{\mathrm{a}}$ & 24.47 (M31)$^{\mathrm{a}}$ & $24.0\pm0.2$\\
\hline
\end{tabular}
\end{center}
$^{\mathrm{a}}$ The low number of stars in these dwarfs makes it difficult to determine their distance modulus, which
is taken as the one of the M31 galaxy. See also section~4 down below.\\
$^{\mathrm{b}}$ The metallicity value for GC is derived from isochrone fitting.
\end{table*}

\subsection{And~XI}
\begin{figure}
\begin{center}
\includegraphics[width=0.75\hsize, angle=270]{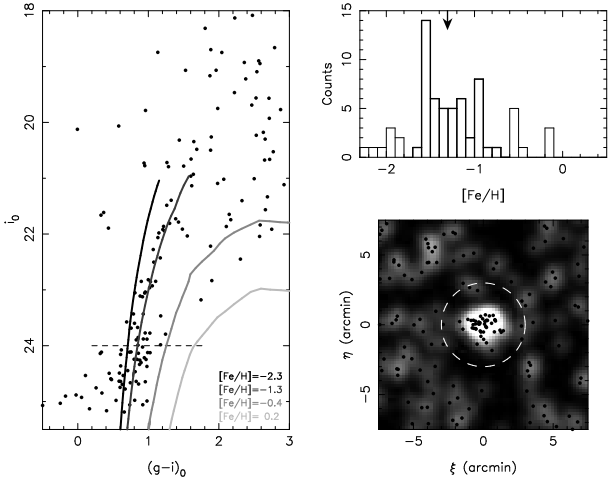}
\caption{\emph{Left:} Color-magnitude diagram of the region within $3'$ of the center of And~XI. Isochrones for a 14~Gyr
population from \citet{girardi04} have been overlaid assuming the distance of M31. The magnitude limit used to
determine the parameters of the dwarf ($i_0<24.0$) is shown as the thin dashed line. Most stars are aligned along a RGB
with a metallicity of $\FeH\sim-1.3$. \emph{Right, top panel:} Metallicity distribution of the corresponding stars.
The distribution is centered around a median value of $-1.3$ (arrow) and only stars with $-1.8<\FeH<-0.8$ are considered
to belong to the dwarf and used to derive its parameters (thick histogram). \emph{Right, bottom panel:} Distribution on
the sky of stars within the And~XI metallicity range. They are clearly well clustered within a $3'$ radius (the white
dashed circle). The background map is obtained by smoothing the star distribution with a $0\mcnd5$ Gaussian, revealing a
rather smooth and elliptical overdensity despite the low number of stars.}
\label{M31dwarf_fig03}
\end{center}
\end{figure}

This satellite is the most luminous of the three new dwarf galaxies found around M31. It is however so faint that the
MegaCam image observed in the $g$ band does not show any visible dwarf, although it does correspond to an overdensity of
stars (bottom right panel of Figure~\ref{M31dwarf_fig03}). The CMD of the region within 3 arcminutes of the central
position of the dwarf is constructed in the left panel of Figure~\ref{M31dwarf_fig03} and shows a RGB, composed of
$\sim40$ stars with an implied metallicity of $\FeH\sim-1.3$. The metallicity distribution of stars with $i_0<24.0$
(top right panel of the same Figure) indeed shows a higher number of stars around the median value of $-1.3$. Only stars
within this peak, that is with $-1.8<\FeH<-0.8$ that populate the RGB (thick histogram) are considered as And~XI stars
and used to determine the parameters of the dwarf. However, the low number of RGB stars makes these determinations quite
uncertain. In particular, the location of the tip of the red giant branch (TRGB), often used to determine the distance
of galaxies, is hard to pinpoint among the contaminating stars from the Galactic foreground and the Andromeda halo. The
absence of a clear horizontal branch in the MegaCam data conspires to prevent a proper determination of the distance
(even though the group of stars at $(g-i)_0\simlt 0.4$ and $i_0\sim25.0$ could represent such a horizontal branch).
Hence, we use the distance of M31 as the distance of And~XI ($m-M=24.47$, \citealt{mcconnachie05}; but see section~4
below).

The center of the dwarf, determined as the mean of the stars within 3 arcminutes and with $-1.8<\FeH<-0.8$ is
$(\alpha,\delta)=(0^{\mathrm{h}}46^{\mathrm{m}}20^{\mathrm{s}},+33\deg48'05'')$. The luminosity profiles of the dwarf in
both the $g$ and $i$ bands are constructed from this center by adding the flux contribution of each star,
corrected from the background contaminating flux that is determined from stars within an annulus between 4 and 10
arcminutes. The light profiles become flat within 2 to 4 arcminutes (confirming that there should not be many stars from the dwarf
that are missed by the $3'$ cut) and yield an average half-light radius of $r_{1/2}=0\mcnd5\pm0\mcnd1$. The uncertainties
are determined from a bootstrap method, randomly re-generating the flux of each star from the flux distribution of
And~XI stars. Since these uncertainties do not take into account the uncertainty on the background flux or the
uncertainties on the metallicity and distance cuts, we double them to yield a more realistic value
$r_{1/2}=0\mcnd5\pm0.2$. At the distance of M31, this converts to $r_{1/2}=115\pm45\pc$.

The MegaCam dataset can also be used to determine a lower limit on the absolute magnitude of the dwarf from the top
three magnitudes of the RGB. Summing up the flux of all And~XI stars yields $M_{g\mathrm{,
min,0}}^{\mathrm{AndXI}}=-4.6$ and $M_{i\mathrm{, min,0}}^{\mathrm{AndXI}}=-5.8$ in the $g$ and $i$ bands respectively.
Using the color equations between MegaCam $(g,i)$ magnitudes and $(V,i)$ INT magnitudes described in \citet{ibata06},
and the INT color equations\footnote{\url{http://www.ast.cam.ac.uk/~wfcsur/technical/photom/colours/}} to derive $(V,I)$
Landolt magnitudes, yields $M_{V\mathrm{, min,0}}^{\mathrm{AndXI}}=-5.0$ in the Landolt V band generally used for such
measures. Although a significant part of the dwarf galaxy luminosity is not taken into account by considering only the
top of the RGB, this value can be normalized by comparison with the MegaCam derived magnitude of And~III with the value
of $M_{V,0}^{\mathrm{AndIII}}=-10.2\pm0.3$ measured by \citet{mcconnachie05} (in doing so, we assume that both 
And~III and And~XI have a similar luminosity function shape). MegaCam observations yield $M_{V\mathrm{,
min,0}}^{\mathrm{AndIII}}=-7.9$, a difference of 2.3 magnitudes compared to the \citet{mcconnachie05} value. The
absolute magnitude of And~XI should therefore approximately be $M_{V,0}^{\mathrm{AndXI}}\simeq M_{{V,0}\mathrm{,
min}}^{\mathrm{AndXI}}-2.3=-7.3$. Obviously uncertainties on this value are significant but a comparison of the CMD of
Figure~\ref{M31dwarf_fig03} with that of And~IX and~X \citep{zucker04,zucker06a} and Ursa Major \citep{willman05} shows
the RGB of And~XI is less populated than that of these two other M31 satellites but more populated than that of UMa.
Hence, And~XI can naively be expected to be less luminous than And~IX and~X that have magnitudes of
$M_{V,0}^{\mathrm{AndIX}}=-8.3\pm0.5$ and $M_{V,0}^{\mathrm{AndX}}=-8.1\pm0.5$ but more luminous than UMa that has
$M_{V,0}^{\mathrm{UMa}}\sim-6.5$. This is indeed what we find, yielding confidence in the measured value and suggesting
an uncertainty of the order of $\pm0.5$~mag.

\subsection{And~XII}
\begin{figure}
\begin{center}
\includegraphics[width=0.75\hsize, angle=270]{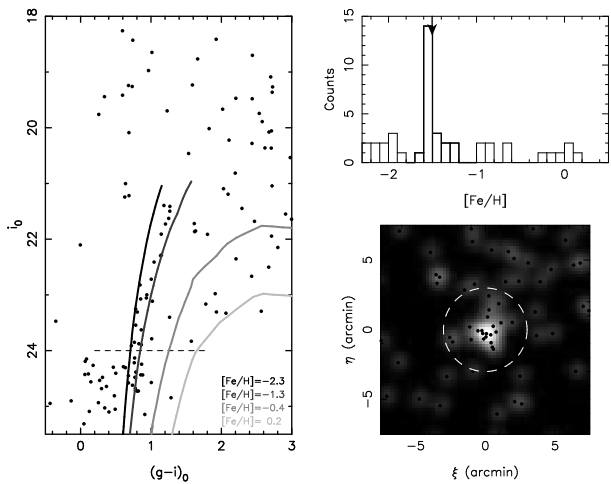}
\caption{As Figure~\ref{M31dwarf_fig03} for the dwarf And~XII.}
\label{M31dwarf_fig04}
\end{center}
\end{figure}

Among the three dwarf galaxies presented in this paper, And~XII is the one with the least populated RGB with only
$\sim20$ stars. The CMD of stars within $3'$ of its central position is shown in the left panel of
Figure~\ref{M31dwarf_fig04} and contains a well defined RGB between the isochrones of metallicities $-2.3$ and $-1.3$. The
metallicity distribution once again contains a strong peak, with a median value of $\FeH=-1.5$. In the following only
stars within the $-1.8<\FeH<-1.0$ metallicity range are considered to belong to the dwarf. The distribution of these
stars on the sky (bottom right panel of the same Figure) shows that, even though there are not many stars in this region
of the sky, they do correspond to an overdensity, centered on
$(\alpha,\delta)=(0^{\mathrm{h}}47^{\mathrm{m}}27^{\mathrm{s}},+34\deg22'29'')$. The probability that a group of stars
are at the same time clustered on the sky and along a RGB is very low (see section~4), strengthening the hypothesis that
And~XII is a dwarf even though it is not directly visible on the MegaCam images.

The derived parameters are also consistent with those of a dwarf, though a very faint one. The half-light radius is
determined as for And~XI from the luminosity profiles that yield $r_{1/2}=0\mcnd55\pm0.2$, corresponding to
$r_{1/2}=125\pm45\pc$ assuming the same distance as M31. The absolute magnitude in the Landolt V band, determined as
before from the MegaCam data, is $M_{V\mathrm{, min,0}}^{\mathrm{AndXII}}=-4.1$ which is normalized to
$M_{V,0}^{\mathrm{AndXII}}=-6.4$. A comparison with UMa ensures this value is sensible since both dwarfs contain a
similar number of stars along their RGB and yield a similar absolute magnitude. However, given the low number of stars,
this value has high uncertainties, of the order of $\pm1.0$.

\subsection{And~XIII}
\begin{figure}
\begin{center}
\includegraphics[width=0.75\hsize, angle=270]{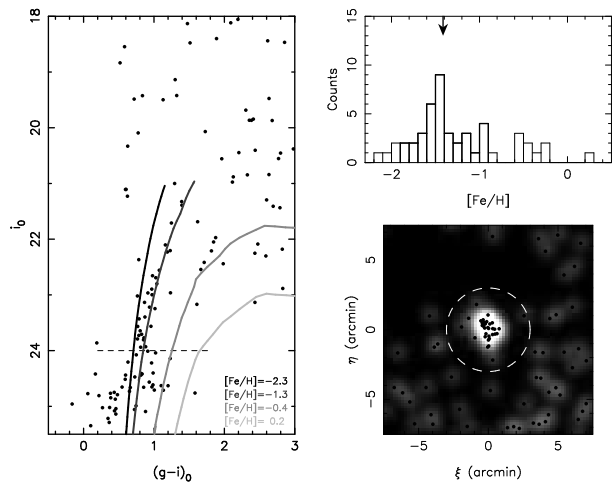}
\caption{As Figure~\ref{M31dwarf_fig03} for the dwarf And~XIII.}
\label{M31dwarf_fig05}
\end{center}
\end{figure}

The third faint dwarf galaxy that appears in the survey is presented in Figure~\ref{M31dwarf_fig05}. Once again, the CMD
shows a RGB populated with $\sim30$~stars within $3'$ of the central position of the dwarf. The metallicity distribution
is also well peaked, with a median value of $\FeH=-1.4$. Since this peak has no clear limits, especially of the low
metallicity side, we select And~XIII stars within $\pm0.5$~dex of the median value. The distribution of these stars on
the sky (bottom right panel of Figure~\ref{M31dwarf_fig05}) reveals a roughly elliptical overdensity. Even if some
contaminants are not taken out of the sample with this metallicity cut, the background correction that is performed
using stars with the same metallicity in an annulus between $4'$ and $10'$ should correct for these. The mean position of
this sample of stars is $(\alpha,\delta)=(0^{\mathrm{h}}51^{\mathrm{m}}51^{\mathrm{s}},+33\deg00'16'')$ and is taken as
the center of the dwarf.

The number of stars in And~XIII is only a little higher than in And~XII, and so the parameters of the dwarf are also
difficult to derive precisely. As above, we use the distance of M31 for the dwarf to account for the difficulty of
determining the tip of the sparsely populated RGB. The half-light radius is determined to be $r_{1/2}=0\mcnd5\pm0.2$,
that is $r_{1/2}=115\pm45\pc$. The absolute magnitude in the V band, determined from the top three magnitudes of the
RGB, is $M_{V\mathrm{, min,0}}^{\mathrm{AndXIII}}=-4.6$ which is normalized as before to
$M_{V,0}^{\mathrm{AndXIII}}=-6.9\pm1.0$.

\begin{figure}
\begin{center}
\includegraphics[width=0.50\hsize, angle=270]{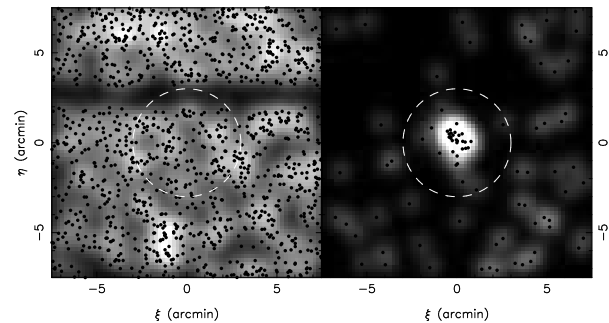}
\caption{Distribution of galaxies (left) and stars (right) in the region around And~XIII. White pixels represent high
density regions. The maps are highly dissimilar (two dimension K-S probability $<10^4$) showing it is improbable that
And~XIII is in fact a misidentified background galaxy cluster. The void at $\eta\sim2\mcnd5$ is due to a gap between
CCDs in the observations.}
\label{M31dwarf_fig06}
\end{center}
\end{figure}

Since the only visible feature of the observed images near And~XIII is a cluster of galaxies, we check that the dwarf is
not artificially created by misidentified background galaxies. It is reassuring that the distribution of galaxies within
a few arcminutes of And~XIII is very different from that of stars (see Figure~\ref{M31dwarf_fig06}). A two dimension
Kolmogorov-Smirnov test \citep{press92} yields a very low probability ($<10^4$) that both maps trace the same population
and given that it is highly improbable that galaxies would align along a RGB, the presence of a background galaxy
cluster close to the center of the dwarf has to be a coincidence and the dwarf is most certainly a genuine association
of stars.

\subsection{GC}
\begin{figure}
\begin{center}
\includegraphics[width=0.9\hsize]{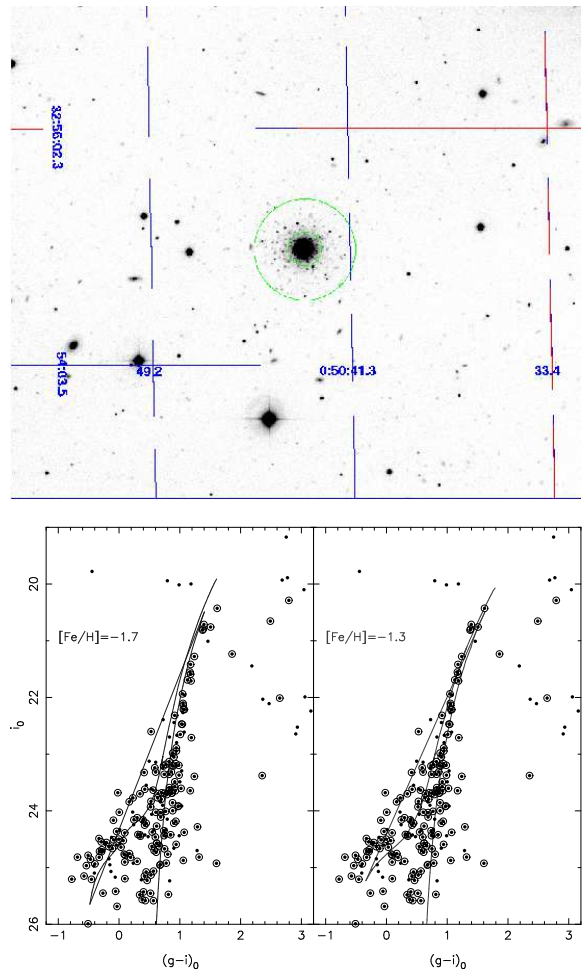}
\includegraphics[width=0.70\hsize, angle=270]{M31dwarf_fig07.ps}
\caption{\emph{Top:} MegaCam image in the $g$ band of the region around the globular cluster. North is to the top and 
east is to the left. \emph{Bottom:} CMD of a region of $1\mcnd5$ around the globular cluster GC. Stars obtained
\emph{via} the Allstar reduction are circled and belong to the central regions of the cluster. The brightest star of the
RGB has a magnitude of $i_0=20.43\pm0.07$. A horizontal branch is also visible from $(g-i,i)_0\sim(-0.6,25.0)$ to
$(g-i,i)_0\sim(0.5,24.0)$ and is used for fitting the isochrones of \citet{girardi04}. The best fits correspond to a
17.8~Gyr populations with $\FeH=-1.7$ (left) and $\FeH=-1.3$ (right).}
\label{M31dwarf_fig07}
\end{center}
\end{figure}

The new globular cluster discovered in the MegaCam survey is located at
$(\alpha,\delta)=(0^{\mathrm{h}}50^{\mathrm{m}}42^{\mathrm{s}}.5,+32\deg54'59\scnd{}6)$. Since the CASU pipeline is not
well suited for separating overlapping stars in crowded regions, the region of the survey around the globular cluster
was re-reduced using the Allstar package \citep{stetson94}. The Allstar reduction is used for the region between 50 and
140 pixels (that is $9''$ to $26''$ of the center of the cluster) with objects with $\chi^2<2.0$ considered to be
stars. The CASU reduction is used over 140~pixels from the cluster to have a galaxy/star separation that is homogeneous
with the rest of the survey (see the top panel of Figure~\ref{M31dwarf_fig07}).

The resulting CMD is presented in Figure~\ref{M31dwarf_fig07} for a region of $1\mcnd5$ around the cluster. Stars recovered
with the Allstar reduction, which are closer to the center, are circled and help populate the RGB. Contrary to the
new dwarf galaxies, the TRGB is easy to determine and the most luminous star of the branch (an Allstar object) appears
to have a magnitude of $i_0=20.43\pm0.07$. Converting this value to the Landolt $I$ band where the TRGB is mainly
invariant with metallicity (see e.g. \citealt{bellazzini01}) yields $I_0=19.94\pm0.1$\footnote{If this star is a
contaminant, the next star along the RGB has $I_0=20.24\pm0.1$, yielding a greater distance of $721\kpc$. However, using
this distance yields a poor isochrone fit to the horizontal branch of the cluster. Hence, we prefer using
$I_0=19.94\pm0.1$ as the magnitude of the tip.}. Assuming a metallicity of $\FeH=-1.3$ for the cluster (see below), the
absolute magnitude of the tip is $M_I=4.05\pm0.1$ \citep{bellazzini01}, we derive a distance modulus of
$(m-M)_0=24.0\pm0.2$. This converts to a distance of $631\pm58\kpc$ and at a projected distance of $\sim100\kpc$ from
M31, that places the cluster at a distance of $175\pm55\kpc$ from the Andromeda galaxy which makes it the most remote
M31 globular cluster currently known, but places it well within the M31 virial radius.

In addition to the well populated RGB, the CMD also shows a horizontal branch from $(g-i,i)_0\sim(-0.6,25.0)$ to
$(g-i,i)_0\sim(0.5,24.0)$. Even though these magnitudes are near the limits of the MegaCam data, the absence of similar
features in the CMDs of Figures~\ref{M31dwarf_fig03}, \ref{M31dwarf_fig04} and \ref{M31dwarf_fig05} that correspond to
larger regions indicates that it is related to the cluster. This horizontal branch, along with the TRGB can be used to
determine the age and metallicity of the cluster from comparison with the \citet{girardi04} isochrones. The best two
fits are shown in the two panels of Figure~\ref{M31dwarf_fig07} and correspond to the oldest isochrones available, with
an age of 17.8~Gyr, and with metallicities of $-1.7$ and $-1.3$ assuming no $\alpha$ enhancement. The fit with
$\FeH=-1.3$ produces the best results and even though an age of 17.8~Gyr is of course improbable, younger isochrones
cannot reproduce the extent of the horizontal branch, meaning the cluster contains a very old population, as is expected
for a globular cluster.

To derive the absolute magnitude of the cluster, we assume most of the luminosity comes from the integrated light in
the inner regions. Hence, a wide aperture is used to derive the magnitude of the cluster as a single object, yielding
background corrected magnitudes of $m_g=16.03$ and $m_i=15.09$. In this region of the sky, the extinction maps of
\citet{schlegel98} give $E(B-V)=0.08$, thus we obtain $m_{g,0}=15.73$ and $m_{i,0}=14.92$. These values convert to
$m_{V,0}=15.5\pm0.2$ where the uncertainties account for the color equations between MegaCam and Landolt magnitudes
and the few stars in the outskirts of the cluster that are not taken into account by this method. Along with the
distance modulus of $(m-M)_0=24.0\pm0.2$ obtained previously, the absolute magnitude of GC is
$M_{V,0}=-8.5\pm0.3$. The half-light radius of the cluster is derived at the same time as $r_{1/2}=0\scnd76$ which
converts to $r_{1/2}=2.3\pm0.2\pc$ at $631\pm58\kpc$. Hence, GC shows typical parameters for a bright globular
cluster (see e.g. Figure~7 of \citealt{huxor04}).

\section{Discussion}
\begin{figure}
\begin{center}
\includegraphics[width=0.60\hsize, angle=270]{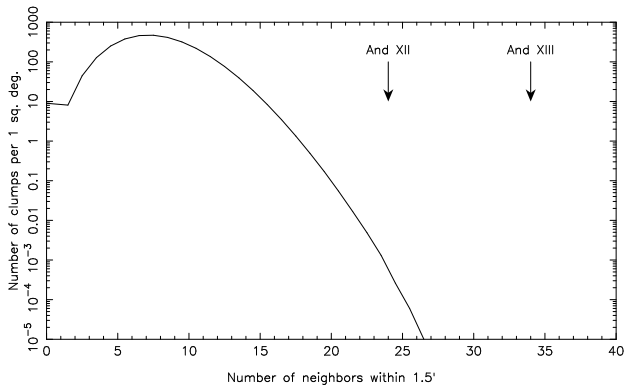}
\caption{Distribution of the number of neighbors within $1\mcnd5$ of each star in a square-degree of artificial
MegaCam survey. $10^5$ square degrees were generated and averaged to obtain this distribution. The arrows represent
the central density of And~XII and~XIII (24 and 34 neighbors within $1\mcnd5$) and have a very low probability to
appear randomly in the survey.}
\label{M31dwarf_fig08}
\end{center}
\end{figure}

Although the globular cluster GC is clearly visible in the MegaCam images, it is not the case for the three dwarf
galaxies so it is natural to wonder if the corresponding overdensities could be only random associations of stars. To
test this possibility, we generate $10^5$ square degrees of artificial survey with the photometric characteristics of 
the 9 MegaCam fields in the outer regions of the M31 halo where no stream and/or structure is apparent. These fields
have an average density of 3050 stars in the CMD region that corresponds to M31 RGB stars. The number of neighbors
within $1\mcnd5$ of each star in the artificial survey is determined and compared to the 24 and 34~stars within the same
radius at the center of And~XII and~XIII. Figure~\ref{M31dwarf_fig08} shows that the smaller of these two overdensities,
And~XII, is expected to be found randomly less than $10^{-3}$ times in a square-degree of the survey. So even in the
whole 57 square-degrees of the MegaCam survey, only $\sim10^{-2}$ such structures are expected. Moreover, this quick
simulation does not take into account the clustering of stars along RGBs which means that randomly finding an
And~XII-like structure in the survey is even less probable. Since And~XI and And~XII contain more stars, we confidently
conclude that the three stellar overdensities are dwarf galaxies and not random associations of stars.

\begin{figure}
\begin{center}
\includegraphics[width=0.60\hsize, angle=270]{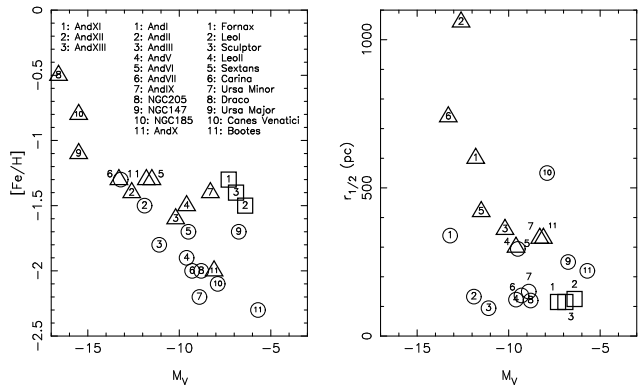}
\caption{Comparison of the parameters of the three new dwarfs (squares), with those of M31 satellites (triangles) and
Milky Way satellites (circles). The correspondence between numbers and galaxies is presented on the left panel. The new
dwarfs seem to be more metal-rich than the other Andromeda satellites, even though the lack of faint dwarfs prevent a
firm conclusion (left). They are also two to three times smaller than the smallest previously-known M31 dwarfs and are
about the size of most Milky Way satellites. (The data are taken from \citealt{mcconnachie05} for most of Andromeda
satellites, And~X values are from \citealt{zucker06a}, UMa values from \citealt{willman05}, CVn values from
\citealt{zucker06b}, Boo values from \citealt{belokurov06} and from \citealt{mateo98} otherwise.)}
\label{M31dwarf_fig09}
\end{center}
\end{figure}

Given their faintness, a detailed comparison of these dwarf galaxies with the other M31 satellites is premature and
requires deeper observations. However, a few interesting points can already be discussed here. Firstly, the three
dwarfs do not seem to extend the $\FeH-M_V$ relation that is followed by the Andromeda satellites
(Figure~\ref{M31dwarf_fig09}, left panel). The lack of other faint M31 satellites as well as the sparse RGB of
the new dwarfs and the different techniques used to derive the metallicities of the various dwarfs make it hard to
ascertain the extent of this offset or if it is due to a flattening of the relation. Nonetheless, at least for the Milky
Way where the faint Boo and CVn dwarfs have recently been found \citep{belokurov06,zucker06b}, fainter dwarfs show more
metal-poor populations. Alternatively, if as proposed by \citet{ricotti05}, these objects correspond to ``true fossil''
galaxies for which star formation was mainly suppressed after reionization, their simulations show that such objects are
expected to have a wide metallicity spread of about 1.0\,dex at the luminosity of And~XI,~XII and~XIII (a few
$10^4\lsun$). This could explain their high metallicity compared to other Andromeda satellites they tag as ``true
fossils'' (e.g. And~I,~II,~III,~V or~VI) but deeper observations are required in order to ascertain this classification
by comparison with the other parameters they determine in their simulations (surface brightness, core radius, central
luminosity density).

A significant difference between the new dwarfs and the other M31 satellites is their small size. Indeed, as
can be seen on the right panel of Figure~\ref{M31dwarf_fig09}, they are two to three times smaller than the smallest
previously-known Andromeda dwarfs. Although the values we derive here and recall in Table~\ref{tableSat} 
have high uncertainties, these cannot account for this difference. A similar difference is observed between Andromeda
and Milky Way satellites and has been proposed by \citet{mcconnachie05} to be due to differences in the formation and/or
evolution of the dwarfs. In particular, \citet{oh95} showed that a dwarf galaxy, of mass $M$ and on an orbit with
semi-major axis $a$, and that lies in the logarithmic potential of its host has a tidal radius $r_t$ that is
proportional to $r_t\propto a\,\Big(\frac{M}{M_h}\Big)^{1/3}$ where $M_h$ is the mass of the host halo within
$a$\footnote{In fact, $r_t$ also depends on the eccentricity of the orbit of the dwarf but this dependency is weak.}.
Even though a precise determination of the tidal radius of the new dwarfs requires deeper observations, the measured
$r_{1/2}$ can be used as an indicator of the size of the objects. Indeed, dwarf galaxies with a small half-light radius
tend to have a small tidal radius (see e.g. Figure~5 of \citealt{mcconnachie05}). If And~XI,~XII and~XIII have been
lying in the M31 halo along with the other Andromeda satellites, their projected distance from M31 ($\sim8\deg$) makes
their semi-major axis at least $\sim110\kpc$ which is of the order of the semi-major axis of the other M31 dwarfs,
implying that the observed difference in $r_{1/2}$ would be related to a difference in mass that should be about 10 to
30 times smaller than the other Andromeda dwarfs. While this could be due to their faintness, it has to be noted that
measurements of the mass of the faint And~IX \citep{chapman05} and UMa \citep{kleyna05} does not make them much less
massive than brighter dwarfs ($\sim10^7\msun$). Alternatively, the small size of And~XI,~XII and~XIII could be the
telltale sign of different formation and/or evolution compared to the other M31 satellites. For instance, they could
have been brought into the M31 halo as satellites of a recently merged or accreted galaxy. This scenario is all the
more interesting since the new dwarfs all lie within only $2\deg$ of one another which could be an indication that 
they form a linked sub-group. The presence of a newly found diffuse stellar stream with $\FeH\sim-1.3$ not far from the
three dwarfs in the M31 halo \citep[see][]{ibata06} also lends weight to this scenario, although the question of the
survival of the dwarfs in such a merger remains open. The radial velocity of the dwarfs is obviously required to
determine if they are linked but their faintness makes them very challenging even for spectrographs on the largest
current telescopes.

\begin{figure}
\begin{center}
\includegraphics[width=0.65\hsize, angle=270]{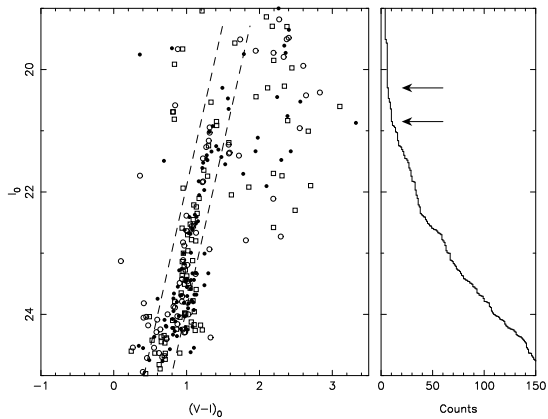}
\caption{Combined CMD of the And~XI (filled circles), And~XII (hollow circles) and And~XIII (squares). The three dwarfs
show strikingly similar RGBs, with a tip between $I=20.3$ and $I=20.85$. The luminosity function of the
RGBs (the region between the two dashed line) is also shown on the right with the two arrows corresponding to the two
determinations of the TRGB. The MegaCam magnitudes have been converted to Landolt magnitudes using the \citet{ibata06}
color equations.}
\label{M31dwarf_fig10}
\end{center}
\end{figure}

If we assume that the galaxies are indeed linked, they should be at about the same distance. Under this assumption 
we sum up their CMDs and measure the RGB tip. This is done on Figure~\ref{M31dwarf_fig10} where the MegaCam 
magnitudes have been converted to Landolt magnitudes. The three RGBs follow strikingly indistinguishable tracks,
suggesting that these three dwarfs, apart from their luminosities, are very similar. The TRGB remains approximative but
it can be determined to be within $I_0=20.30$ and $I_0=20.85$ depending on whether the four stars (3 And~XI stars and one
And~XIII star) between the two values are considered to belong to the dwarfs or not. Assuming $M_I=-4.05$ as before,
this yields a distance of 740 to 955\kpc. In any case, this puts the dwarfs well within the virial radius of M31.
Fitting the \citet{girardi04} isochrones with the two values derived for the tip yields metallicities of $-1.3$ and
$-1.7$ respectively. The uncertainty on the distance modulus ($\pm0.3$) will of course add to the uncertainties on the
absolute magnitude and half-light radius of the dwarfs but that will not change the general conclusions presented here.

Finally, the panoramic view of the MegaCam survey can be used to provide some constraints on the missing satellite
problem. Five dwarf galaxies with luminosity $M_V\simlt-6.4$ are found in the survey (And~II, And~III, And~XI, And~XII
and And~XIII). Assuming a Poissonian distribution, the survey would therefore contains $5\pm2$ satellites within the
given magnitude limit. Moreover, the integration of a NFW profile along the line of sight using suitable parameters for
M31 (concentration of $c=10$ and virial radius of 313\kpc; \citealt{klypin02}) shows that the survey covers 11\% of the
total M31 halo. Since satellites in the central regions of the halo tend to be harassed by tidal interaction and are
more easily destroyed, the maximum number of satellites of the Andromeda galaxy with $M_V\simlt-6.4$ can therefore be 
estimated to be $45\pm20$. Even if we assume these objects reside in massive dark matter halos of $\sim10^7\msun$ as
has been measured for Ursa Major \citep{kleyna05}, And~IX \citep{chapman05} and Boo \citep{munoz06}, this number is
still one order of magnitude smaller than the number of CDM satellites \citep[e.g.][]{stoehr02}. Thus, it seems faint
dwarf galaxies cannot alone account for the so-called missing satellites except if their currently measured masses are
underestimated by a factor of $\sim10$, for instance if they have dark matter halos that extend beyond their luminous
counterpart. It has to be noted however, that larger and more diffuse structures could be missed by the visual
inspection that led to the discovery of the new dwarfs. A search for these sub-structures is under way and will be
presented in an up-coming paper (Martin et al., in prep.).

\section{Conclusions}
The inspection of the 57 square-degrees of our MegaCam survey of a quadrant of the M31 halo, from a projected
distance of 50\kpc\ to 150\kpc, reveals the presence of four new satellites of the Andromeda galaxy down to $M_V=-6.4$.
One of these is the farthest M31 globular cluster currently known. The other three satellites are faint dwarf galaxies
that all lie within two degrees of one another, have a similar metallicity and are two to three times smaller than the
smallest of previously-known M31 satellite galaxies. All these characteristics may indicate that they are linked
together, have a common origin and were brought in the M31 halo by the same mechanism. Radial velocity measurements are
necessary to confirm that they are kinematically associated and could also be used to determine if they exhibit the same
high mass-to-light ratio that is found for the faint And~IX \citep{chapman05}, UMa \citep{kleyna05} and Boo
\citep{munoz06}. The sparsely populated RGBs of these new dwarfs will however prove a challenge for current telescopes.
Deeper photometric observations are also required to better characterize the new structures, constrain their parameters
and their stellar populations and search for tidal features in their outskirts. Their relation to the asymmetric
distribution of M31 satellites \citep{mcconnachie06} would also be very interesting to study.

The MegaCam survey also gives the first opportunity to study a significant portion of a galactic halo in search of
substructures without any of the bias, such as projection effects and foreground contamination, that plague Milky Way
studies. From the surveyed area, we estimate that a maximum of $45\pm20$ satellites with $M_V\simlt-6.4$ orbit around
the Andromeda galaxy, which is one order of magnitude smaller than the number of substructures expected from CDM
models, except if they reside in more massive halos than currently measured. The missing satellites thus do not seem to
be highly dark matter dominated dwarf galaxies unless many are even fainter or more extended than those found here.

\section*{Acknowledgments}
We wish to thank the CFHT staff for performing the MegaCam observations in queue mode.

\newcommand{\mnras}{MNRAS}
\newcommand{\pasa}{PASA}
\newcommand{\nat}{Nature}
\newcommand{\araa}{ARAA}
\newcommand{\aj}{AJ}
\newcommand{\apj}{ApJ}
\newcommand{\apjl}{ApJ}
\newcommand{\apjs}{ApJSupp}
\newcommand{\aap}{A\&A}
\newcommand{\aaps}{A\&ASupp}
\newcommand{\pasp}{PASP}


\begin{thebibliography}{}
%
\bibitem[Armandroff, Davies \& Jacoby(1998)]{armandroff98}
        Armandroff T. E., Davies J. E. \& Jacoby G. H. 1998, \aj\ 116, 2287
%
\bibitem[Armandroff, Jacoby \& Davies(1999)]{armandroff99}
        Armandroff T. E., Jacoby G. H. \& Davies J. E. 1999, \aj\ 118, 1220
%
\bibitem[Bellazzini, Ferraro \& Pancino(2001)]{bellazzini01}
        Bellazzini M., Ferraro F. R. \& Pancino E. 2001, \apj\ 556, 635
%
\bibitem[Belokurov et al.(2006)]{belokurov06}
        Belokurov et al. 2006, \apj\ submitted, astro-ph/0604355
%
\bibitem[Bullock, Kravstov \& Weinberg(2000)]{bullock00}
        Bullock J. S., Kravstov A. V. \& Weinberg D. H. 2000, \apj\ 539, 517
%
\bibitem[Chapman et al.(2005)]{chapman05}
        Chapman S., Ibata R., Ferguson A. M. N., Lewis G., Irwin M. \& Tanvir N. 2005, \apj\ submitted
%
\bibitem[Ferguson et al.(2002)]{ferguson02}
        Ferguson A. M. N., Irwin M. J., Ibata R. A., Lewis G. F. \& Tanvir N. R. 2002, \aj\ 124, 1452
%
\bibitem[Girardi et al.(2004)]{girardi04}
        Girardi L., Grebel E. K. Odenkirchen M. \& Choisi C. 2004, \aap\ 422, 205
%
\bibitem[Grebel(2001)]{grebel01}
        Grebel E. K. 2001, Ap\&SS 277, 231
%
\bibitem[Harbeck et al.(2005)]{harbeck05}
        Harbeck D., Gallagher J. S., Grebel E. K., Koch A. \& Zucker D. B. 2005, \apj\ 623, 159
%
\bibitem[Huxor et al.(2004)]{huxor04}
        Huxor A. P., Tanvir N. R., Irwin M. J., Ibata R. A., Collett J. L., Ferguson A. M. N., Bridges T. \& Lewis G.
F. 2004, \mnras\ 360, 1007
%
\bibitem[Ibata et al.(2001)]{ibata01}
        Ibata R. A., Irwin M. J., Lewis G. F., Ferguson A. M. N. \& Tanvir N. R. 2001, \nat\ 412, 49
%
\bibitem[Ibata et al.(2005)]{ibata05}
        Ibata R. A., Chapman S. C., Ferguson A. M. N., Lewis G. F., Irwin M. J. \& Tanvir N. R. 2005, \apj\ 634, 287
%
\bibitem[Ibata et al.(2006)]{ibata06}
        Ibata R. A., Martin N. F., Chapman S. C., Ferguson A. M. N., Irwin M. J., Lewis G. F. \& Tanvir N. R. 2006,
\mnras\ submitted
%
\bibitem[Irwin \& Lewis(2001)]{irwin01}
        Irwin M. \& Lewis J. 2001, New Astronomy Review, 45, 105
%
\bibitem[Kleyna et al.(2005)]{kleyna05}
        Kleyna J. T., Wilkinson M. I., Evans N. W. \& Gilmore G. 2005, \apj\ 630, L141
%
\bibitem[Klypin et al.(1999)]{klypin99}
        Klypin A., Kravstov A. V., Velenzuela O. \& Prada F. 1999, \apj\ 522, 82
%
\bibitem[Klypin, Zhao \& Sommerville(2002)]{klypin02}
        Klypin A., Zhao H. \& Sommerville R. S. 2002, \apj\ 573, 597
%
\bibitem[Mackey \& Gilmore(2004)]{mackey04}
        Mackey A. D. \& Gilmore G. 2004, \mnras\ 355, 504
%
\bibitem[Mateo(1998)]{mateo98}
        Mateo M. 1998, \araa\ 36, 435
%
\bibitem[McConnachie et al.(2004)]{mcconnachie04}
        McConnachie A. W., Irwin M. J., Lewis G. F., Ibata R. A., Chapman S. C., Ferguson A. M. N. \& Tanvir N. 2004,
\mnras\ 351, L94
%
\bibitem[McConnachie et al.(2005)]{mcconnachie05}
        McConnachie A. W., Irwin M. J., Ferguson A. M. N., Ibata R. A., Lewis G. F. \& Tanvir N. 2005, \mnras\ 356, 979
%
\bibitem[McConnachie \& Irwin(2006)]{mcconnachie06}
        McConnachie A. W. \& Irwin M. J. 2006, \mnras\ 365, 902
%
\bibitem[Moore et al.(1999a)]{moore99a}
        Moore B., Ghigna S., Governato F., Lake G., Quinn T., Stadel J. \& Tozzi P. 1999a, \apj\ 524, L19
%
\bibitem[Moore et al.(1999b)]{moore99b}
        Moore B., Quinn T., Governato F., Stadel J. \& Lake G. 1999b, \mnras\ 310, 1147
%
\bibitem[Moore et al.(2001)]{moore01}
        Moore B., Calc\'aneo-Rold\'an C., Stadel J., Quinn T., Lake G., Ghigna S. \& Governato F. 2001, Phys Rev D\
64, 3508
%
\bibitem[Mu\~noz et al.(2006)]{munoz06}
        Mu\~noz R. R., Carlin J. L., Frinchaboy P. M., Nidever D. L., Majewski S. R. \& Patterson R. J. 2006, \apj\
submitted, astro-ph/0606271
%
\bibitem[Navarro, Frenk \& White(1997)]{navarro97}
        Navarro J. F., Frenk C. S. \& White S. D. M. 1997, \apj\ 490, 493
%
\bibitem[Oh, Lin \& Aarseth(1995)]{oh95}
        Oh K. S., Lin D. N. C. \& Aarseth S. J. 1995, \apj\ 442, 142
%
\bibitem[Press et al.(1992)]{press92}
        Press W. H., Flannery B. P., Teukolsky S. A., Vetterling W. T. 1992, Numerical Recipes. Cambridge Univ. Press,
Cambridge
%
\bibitem[Ricotti \& Gnedin(2005)]{ricotti05}
        Ricotti M. \& Gnedin N. 2005, \apj\ 629, 259
%
\bibitem[Schlegel, Finkbeiner \& Davis(1998)]{schlegel98}
        Schlegel D., Finkbeiner D. \& Davis M. 1998, \apj\ 500, 525
%
\bibitem[Sommerville(2002)]{sommerville02}
        Sommerville R. S. 2002, \apj\ 572, L23
%
\bibitem[Stetson(1994)]{stetson94}
        Stetson P. B. 1994, \pasp\ 106, 250
%
\bibitem[Stoehr et al.(2002)]{stoehr02}
        Stoehr F., White S. D. M., Tormen G. \& Springel V. 2002, \mnras\ 335, L84
%
\bibitem[Tully et al.(2002)]{tully02}
        Tully R. B., Sommerville R. S., Trentham N. and Verheijen M. A. W. 2002, \apj\ 569, 573
%
\bibitem[Willman et al.(2005)]{willman05}
        Willman et al. 2005, \apj\ 626, L85
%
\bibitem[Zucker et al.(2004)]{zucker04}
        Zucker et al. 2004, \apj\ 612, L121
%
\bibitem[Zucker et al.(2006a)]{zucker06a}
        Zucker et al. 2006a, \apj\ submitted, astro-ph/0601599
%
\bibitem[Zucker et al.(2006b)]{zucker06b}
        Zucker et al. 2006b, \apj\ 643, L103
%
\end{thebibliography}
\end{document}